\documentclass[12pt,a4paper]{article}

\usepackage{epsfig}
\usepackage{amsmath,amsfonts, amssymb}

\begin{document}
\bibliographystyle{plain}
\pagenumbering{arabic}
\newcommand{\tr}[1]{\mbox{Tr} \, #1 }
\newtheorem{theorem}{Theorem}
\newtheorem{lemma}[theorem]{Lemma}
\newtheorem{prop}[theorem]{Proposition}
\newtheorem{example}[theorem]{Example}
\def \iden {{\bf{I}}}
\newcommand \s {\widetilde{s}}

\renewcommand{\theequation}{\arabic{equation}}
\renewcommand{\thetheorem}{\arabic{theorem}}
\renewcommand{\thesection}{\arabic{section}}

\newenvironment{proof}
{\noindent{\bf Proof\ }}{{\hfill \endbox }\par\vskip2\parsep}

\newcommand{\conpr}{\buildrel{\pr}\over\longrightarrow}
\newcommand{{\Tr}}{\rm Tr}

\newcommand{\RR}{{\mathbb R}}
\newcommand{\pr}{{\mathbb P}}
\newcommand{\ZZ}{{\mathbb Z}}
\newcommand{\NN}{{\mathbb N}}
\newcommand{\CC}{{\mathbb C}}

\newcommand{\hilb}{{\cal{H}}^{\otimes n}}

\newcommand{\EE}{{\mathbb E}}
\newcommand{\ov}[1]{\overline{#1}}

\parindent=0pt \parskip=10pt

\setcounter{equation}{0}

\newcommand \qed {\vrule height5pt width5pt}
\newcommand{\be}{\begin{equation}}
\newcommand{\ee}{\end{equation}}
\newcommand{\bea}{\begin{eqnarray}}
\newcommand{\eea}{\end{eqnarray}}
\newcommand \boundary {\partial}
\newcommand \eps {\epsilon}
\newcommand \half {\frac{1}{2}}
\newcommand \one {{\mbox{\bf 1}}}
\newcommand{\restrict}{{\upharpoonright}}
\newcommand \x {{\underline{x}}}
\def\reff#1{(\ref{#1})}
%

\font\titlefnt=cmr6 scaled \magstep3

\title{\bf{Multiplicativity of Maximal $p$--Norms in \\Werner--Holevo Channels for $1 < p \le 2$}}

\author{Nilanjana Datta 
\\Statistical Laboratory
\\Centre for Mathematical Sciences
\\University of Cambridge
\\Wilberforce Road, Cambridge CB30WB
\\ email: n.datta@statslab.cam.ac.uk
}

\maketitle
\begin{abstract}
Recently, King and Ruskai \cite{kingrus} conjectured  
that the maximal $p$--norm of the Werner--Holevo channel 
is multiplicative for all $1\le p \le 2$.
In this paper we prove this conjecture. 
Our proof relies on certain convexity and monotonicity 
properties of the $p$--norm.
\end{abstract}
\section{Introduction}
A quantum channel $\Phi$ is described by a completely positive
trace--preserving map \cite{nc, kraus}, which acts on an input density matrix $\rho$
to yield the output $\Phi(\rho)$. Under the effect of 
noise present in the channel, pure input states are typically 
transformed into mixed output states. The amount of noise present
in the channel can be estimated by determining how close the
output $\Phi(\rho)$ is to a pure state when the input $\rho$ 
is pure. In other words, the output purity provides a 
yardstick for the level of noise in the channel. There are various
measures of output purity, one of them being the maximal $p$--norm 
of the channel. It is defined as follows
\be
\nu_p(\Phi) := \sup_{\rho} \left\{ ||\Phi(\rho)||_p\right\},
\label{nup}
\ee
where ${\displaystyle{||\Phi(\rho)||_p = 
\Bigl[\tr\left(\Phi(\rho)\right)^p \Bigr]^{1/p}}}$ is the $p$--norm
of $\Phi(\rho)$ and $1\le p\le \infty$. 
The case $p=\infty$ corresponds to the operator norm. 
In \reff{nup} the supremum is taken over all input density matrices.
However, due to convexity of the map $\rho \mapsto ||\Phi(\rho)||_p $, 
it suffices to restrict this supremum to pure states. It is clear that
$||\Phi(\rho)||_p \le 1$, since $\Phi(\rho)$ is a density matrix. The equality
holds if and only if the latter is a pure state. Hence  
$\nu_p(\Phi) \le 1$ with equality if and only if there is a pure state
$\rho$ for which the output $\Phi(\rho)$ is also pure. Thus $\nu_p(\Phi)$
provides a measure of the maximal purity of outputs from a quantum channel
$\Phi$. 

The maximal $p$--norms of two quantum channels
$\Phi$ and $\Psi$ are said to be multiplicative if 
\be
\nu_p(\Phi \otimes \Psi) = \nu_p(\Phi).\nu_p(\Psi).
\label{mult1}
\ee
This multiplicativity was conjectured by Amosov, Holevo and Werner in
\cite{AHW}. In the limit $p \rightarrow 1$, \reff{mult1} implies the
additivity of another natural measure of the output purity, namely the von 
Neumann entropy. The multiplicativity \reff{mult1} has been proved explicitly for various
cases. For example, it is valid for all integer values of $p$, when $\Phi$ and $\Psi$ are 
tensor products of depolarizing channels \cite{AH}. It also holds   
for all $p\ge 1$ when $\Psi$ is an 
arbitrary quantum channel and $\Phi$ is any one of the following: 
$(i)$ an entanglement breaking channel \cite{kingent}, $(ii)$ a unital qubit 
channel \cite{kingqubit} or $(iii)$ a depolarizing channel in any dimension 
\cite{kingdep}. However, it is now
known that the conjecture is not true in general. A counterexample
to the conjecture was given in \cite{HW}, for $p > 4.79$ in the case in 
which $\Phi$ and $\Psi$ are Werner--Holevo
channels, defined by \reff{channel}.  

The  Werner--Holevo channel $\Phi_d$ of dimension $d < \infty$ 
is defined by its action on any complex $d \times d$ 
matrix $\mu$ as follows
\begin{equation}
\Phi_d (\mu)=\frac{1}{d-1}\bigl({\iden}\,\mathrm{Tr}(\mu )-\mu ^{T}\bigr).
\label{channel}
\end{equation} 
Here $\mu^{T}$ denotes the transpose of $\mu $, and ${\iden}$ is
the $d\times d$ unit matrix. In particular, the action of the channel
on any density matrix $\rho$ is given by
\begin{equation}
\Phi_d (\rho )=\frac{1}{d-1}\bigl({\iden}-\rho ^{T}\bigr).
\label{channel2}
\end{equation}
Werner and Holevo \cite{HW} proved that the conjecture \reff{mult1} 
was false for $p>4.79$
when $\Phi=\Psi= \Phi_d$ with $d=3$. The validity of the conjecture 
for smaller values of $p$ for this channel
was an open question. Recently it was proved \cite{kingrus, af} 
that \reff{mult1} is true for $p=2$ for 
the above channel $\Phi_d$ with $d \ge 2$. In fact, 
in \cite{kingrus} the multiplicativity \reff{mult1} was proved in 
a more general setting, namely one in which $\Phi$ is a Werner--Holevo channel
but $\Psi$ is any arbitrary channel. Moreover, in \cite{kingrus} 
the multiplicativity \reff{mult1} was conjectured to hold for all 
$1\le p \le 2$ for Werner--Holevo channels.  
This paper provides a proof of this conjecture.

The precise statement of our result is given in Theorem 1 of 
Section \ref{main}.
We would like to note that while writing our results, we were made aware
of an almost simultaneous but independent and alternative proof of the
conjecture put forth in \cite{kingrus}. This is contained in the recently 
posted body of work in \cite{af}. However, not only do we present an 
alternative approach
to the same conjecture, but this paper also provides the result encapsulated
in Lemma \ref{lem}, which would be of independent interest. 
\section{Main result}
\label{main}
 Following the discussion in the Introduction, we write the maximal $p$--norms 
for a single Werner--Holevo channel $\Phi_d$ and the product channel
$\Phi_{d_1} \otimes \Phi_{d_2}$ as

\bea
\nu_p(\Phi_d) &=& 
\max_{|\psi\rangle \in {\cal{H}} \atop{ ||\psi ||=1}} \left\{ ||\Phi_d(|\psi\rangle
\langle \psi|)||_p\right\},\nonumber\\
{\hbox{and      }} \quad \nu_p(\Phi_{d1} \otimes \Phi_{d2}) &=& \max_{|\psi_{12}\rangle \in {
\mathcal{H}}_1 \otimes {\mathcal{H}}_2 \atop{ ||\psi_{12} ||=1}}\,
 \left\{ ||\Phi_{1}\otimes \Phi_{2})(|\psi_{12}\rangle\langle \psi_{12}|
||_p\right\},
\label{mult22}
\eea
respectively. In the above, ${\mathcal{H}} \simeq {\mathbf{C}}^{d}$ and 
${\mathcal{H}}_{i}\simeq {\mathbf{C}}^{d_i}$ for $i=1,2$.
Our main result is stated in the following theorem.
\begin{theorem}
\label{theo}
Let $\Phi_d$ denote a Werner--Holevo channel of dimension $d$. Then
the multiplicativity of the maximal $p$--norms
\be
\nu_p(\Phi_{d_1} \otimes \Phi_{d_2}) = \nu_p(\Phi_{d_1}).\nu_p(\Phi_{d_2}),
\label{mult2}
\ee
holds for all $1 \le p \le 2$, for arbitrary dimensions $d_1, d_2 \ge 2$.  
\end{theorem}

To prove Theorem 1 we will make use of the method developed in \cite{suff}
and of certain results proved in it. It is useful to consider the 
Schmidt decomposition of $|\psi _{12}\rangle$
\begin{equation}
|\psi _{12}\rangle =\sum_{\alpha =1}^{d}\sqrt{\lambda _{\alpha }}|\alpha
;1\rangle \otimes |\alpha ;2\rangle .  \label{schmidt}
\end{equation}
Here ${d}=\min \bigl[d_{1},d_{2}\bigr]$ and 
$\left\{ |\alpha ;j\rangle \right\} $ is an orthonormal basis in ${\
\mathcal{H}}_{i}$, $i=1,2$. 
The Schmidt coefficients $\lambda_\alpha$, $\alpha=1,2,\ldots,d$, and 
hence also the vector of Schmidt coefficients ${\underline{\lambda }}:=
(\lambda _{1},\ldots
 ,\lambda _{d})$, 
vary in the $({d}-1)-$dimensional
simplex $\Sigma _{d}$, defined by the constraints
\begin{equation}
\lambda _{\alpha }\geq 0\quad ;\quad \sum_{\alpha =1}^{d}\lambda _{\alpha
}=1.  \label{cons}
\end{equation}
Note that the vertices of $\Sigma_d$ correspond to unentangled vectors
$|\psi_{12}\rangle = |\psi_{1}\rangle \otimes |\psi_{2}\rangle$.
To prove Theorem 1 it is sufficient to show that the maximum on the RHS
of \reff{mult2} is achieved for unentangled vectors. In other words, 
we need to prove that this maximum occurs at the vertices of the simplex
$\Sigma_d$.

Using the Schmidt decomposition \reff{schmidt}, the input to the product channel can be expressed as
\begin{equation}
|\psi _{12}\rangle \langle \psi _{12}| =
\sum_{\alpha ,\beta=1}^{d}
\sqrt{\lambda _{\alpha }\lambda _{\beta }}|\alpha ;1\rangle
\langle \beta ;1|\otimes |\alpha ;2\rangle \langle \beta ;2|.
\label{matrix}
\end{equation}
The output of the channel is the density matrix given by
\be
\sigma_{12}({\underline{\lambda }}):=\left( \Phi _{1}\otimes \Phi _{2}\right)
\left( |\psi _{12}\rangle \langle \psi _{12}|\right) =\sum_{\alpha ,\beta
=1}^{d}\sqrt{\lambda _{\alpha }\lambda _{\beta }}\Phi _{1}(|\alpha ;1\rangle
\langle \beta ;1|)\otimes \Phi _{2}(|\alpha ;2\rangle \langle \beta ;2|).
\label{matrix2}
\ee
 
We prove Theorem 1 by showing that 
$$\left[\nu_p(\Phi_{d_1} \otimes \Phi_{d_2})\right]^p = 
\left[\nu_p(\Phi_{d_1})\right]^p \,\left[\nu_p(\Phi_{d_2})\right]^p 
\quad {\hbox{ for any }} d_1, d_2 \ge 2 \quad {\hbox{and }} 1<p\le 2.$$
The multiplicativity  \reff{mult2} holds trivially for $p=1$ since 
$\tr {\widetilde{\rho}}=1$ for any density matrix ${\widetilde{\rho}}$.
Note that
\bea
\left[\nu_p(\Phi_{d_1} \otimes \Phi_{d_2})\right]^p 
&=& \max_{|\psi_{12}\rangle \in {
\mathcal{H}}_1 \otimes {\mathcal{H}}_2 \atop{ ||\psi_{12} ||=1}}\,
 \left\{ ||(\Phi_{1}\otimes \Phi_{2})(|\psi_{12}\rangle
\langle \psi_{12}|)||_p^p
\right\}\nonumber\\
&=& \max_{{\underline{\lambda }} \in \Sigma_d} \sum_{i=1}^{d_1d_2} 
\left(E_i({\underline{\lambda }})\right)^p,
\eea
where $\{E_i({\underline{\lambda }}), i =1,2, \ldots, d_1d_2\}$ denotes
the set of eigenvalues of the channel output 
$\sigma_{12}({\underline{\lambda }})$. These eigenvalues were studied
in detail in \cite{suff} and were found to be divided into the three
classes  given below. Here we assume for definiteness that 
$d_{1}\leq d_{2},$ so that $d=d_{1}.$
\begin{enumerate}
\item{There are $d(d-1)$ eigenvalues given by
$$
e_{\alpha \beta }:=\frac{1}{(d_{1}-1)(d_{2}-1)}\bigl(1-\lambda _{\alpha
}-\lambda _{\beta }\bigr),\quad \alpha \neq \beta ,\,\,\alpha ,\beta
=1,2,\ldots ,{d}.
$$
}
\item{There are $d$ eigenvalues given by 
$$
h_{\alpha }:=\frac{(1-\lambda _{\alpha })}{(d_{1}-1)(d_{2}-1)};\quad \alpha
=1,2,\ldots ,{d},
$$
each of multiplicity ${d}_{2}-{d}_{1}$.}
\item{There are ${d}$ eigenvalues of the form
\be
g_{\alpha }:=\frac{\gamma _{\alpha }}{(d_{1}-1)(d_{2}-1)},\quad \alpha
=1,2,\ldots ,{d},
\label{gee1}
\ee
where $\gamma _{\alpha }$ are the roots of the equation
\begin{equation}
\prod_{\alpha =1}^{d}(1-2\lambda _{\alpha }-\gamma )\left\{ 1+\sum_{\alpha
^{\prime }=1}^{d}\frac{\lambda _{\alpha ^{\prime }}}{(1-2\lambda _{\alpha
^{\prime }}-\gamma )}\right\} =0.  \label{eigen2}
\end{equation}
}
\end{enumerate}
Using the constraint \reff{cons} we find that
$$
\sum_{1\leq \alpha ,\beta \leq {d}\atop{\alpha \neq \beta}}e_{\alpha
\beta }=\frac{d_{1}-2}{d_{2}-1};\quad \sum_{\alpha =1}^{d}h_{\alpha }=\frac{1
}{d_{2}-1},
$$
and using the fact that the sum of all eigenvalues 
of $\sigma_{12}({\underline{\lambda }})$ is equal to $1$ we get
$$
\sum_{\alpha =1}^{d}g_{\alpha }=\frac{d_{2}-d_{1}}{d_{2}-1}.
$$
These relations allow us to define the following sets of non--negative
variables
\bea
{\widetilde{e_{\alpha\beta}}} &:=& \left(\frac{d_{2}-1}{d_{1}-2}\right)e_{\alpha\beta}
\quad \alpha, \beta = 1,2 \ldots, d; \, \alpha \ne \beta,\\ 
{\widetilde{h_{\alpha}}} &:=&\left({d_{2}-1}\right)h_{\alpha}\quad 
\alpha = 1,2 \ldots, d, \\
{\widetilde{g_{\alpha}}} &:=&\left(\frac{d_{2}-1}{d_{2}-d_{1}}\right)g_{\alpha} \quad 
\alpha = 1,2 \ldots, d, \label{gee2} 
\eea
such that the sum of each of these sets of variables is equal to unity, i.e.
$$\sum_{1\leq \alpha ,\beta \leq {d}\atop{\alpha \neq \beta}}
{\widetilde{e_{\alpha\beta}}}=1 \,\,;\,\,
\sum_{\alpha =1}^{d}{\widetilde{h_{\alpha}}}=1 \,\,;\,\,
\sum_{\alpha =1}^{d} {\widetilde{g_{\alpha}}} =1.$$
Hence, we can write
\bea
||\sigma_{12}({\underline{\lambda }})||_p^p
&=& \left(E_i({\underline{\lambda }})\right)^p\nonumber\\
&=& c_1 \sum_{1\leq \alpha ,\beta \leq {d}\atop{\alpha \neq \beta}}
{\widetilde{e_{\alpha
\beta }}}^p + c_2 \sum_{\alpha=1}^d {\widetilde{h_{\alpha}}}^p + c_3 
\sum_{\alpha=1}^d 
{\widetilde{g_{\alpha}}}^p
\nonumber\\
&:=& T_1({\underline{\lambda}}) + T_2({\underline{\lambda}})   
+ T_3({\underline{\lambda}}),
\eea
where $c_1, c_2$ and $c_3$ are constants depending on the 
dimensions $d_1$ and $d_2$.

The function $f(x) := x^p$, where $1 <p \le 2$ is convex for
$ x \ge 0$. Hence, $T_1({\underline{\lambda}})$ is a convex function
of the variables ${\widetilde{e_{\alpha \beta }}}$. These variables are
affine functions of the Schmidt coefficients $\lambda_1, \ldots, \lambda_d$.
Hence, $T_1({\underline{\lambda}})$ is a convex function of 
$\ {\underline{\lambda }}$ and attains its global 
maximum at the vertices of the simplex $\Sigma _{d}$. The same argument 
applies to $T_2({\underline{\lambda}})$ since the
variables ${\widetilde{h_{\alpha}}}$ are also affine functions of the 
Schmidt coefficients. The function 
$T_3({\underline{\lambda}})$ is however not necessarily a convex function of 
$\ {\underline{\lambda }}$. In spite of this, it too achieves its maximum 
value at the vertices of $\Sigma_d$. This follows from the following theorem.
\begin{theorem}
\label{theo2}
The function $T_3({\underline{\lambda}})$ is Schur-convex in $\underline{ \lambda }\in \Sigma_{d}
$ i.e., $\underline{\lambda }\prec \underline{\lambda }^{\prime }\, \implies
T_3\left( {\ \underline{\lambda }}\right) \leq T_3\left( {\underline{
\lambda }}^{\prime }\right)$, where $\prec $ {denotes the stochastic 
majorization}
(see the Appendix).
\end{theorem}

Since every $\underline{\lambda }\in \Sigma _{d}$ is majorized by the
vertices of $\Sigma _{d}$, Theorem 2 implies 
that $T_3(\underline{\lambda })$
also attains its maximum at the vertices. Thus 
$||\sigma_{12}({\underline{\lambda}})||_p^p = T_1(\underline{\lambda })
+ T_2(\underline{\lambda }) + T_3(\underline{\lambda })$ 
is maximized at the vertices of $\Sigma _{d}$. As was observed, this implies the 
multiplicativity (\ref{mult2}).

To prove Theorem \ref{theo2} we use the following lemma, which is proved in Section
\ref{proof}. 

%

\begin{lemma} 
\label{lem}
Let $f(\x) := \sum_{i=1}^n x_i^p$, where $\x = (x_1, x_2,\ldots, 
x_n)$ with each $x_i \ge 0$ and $\sum_{i=1}^n x_i = 1$. For $1 <p <2$, 
$f(\x)$ is a monotonically non--increasing function of  
the elementary symmetric polynomials $s_{q}(x_{1},x_{2},\ldots ,x_{n})$ 
for $2 \le q \le n$, where 
\bea
s_{k}(x_{1},x_{2},\ldots ,x_{n}) &:=& \sum_{1\leq i_{1}<i_{2}\cdots 
<i_{k}\leq n}x_{i_{1}}x_{i_{2}}\ldots x_{i_{k}}\quad {\hbox{for  }} 
k=1,2,3,\ldots,n.
\label{sym}
\eea
\end{lemma}

Note that $T_3(\underline{\lambda}):= c_3 \sum_{\alpha=1}^d 
{\widetilde{g_{\alpha}}}^p,$ where  
$${\widetilde{g_{\alpha}}} \ge 0 
\quad {\hbox{ and}} \quad   \sum_{\alpha=1}^d {\widetilde{g_{\alpha}}} =1.$$
The variables 
${\widetilde{g_{\alpha}}}$
are proportional to the roots $\gamma_\alpha$ of eq.\reff{eigen2} (see \reff{gee1} and \reff{gee2}),
which are obviously functions of the Schmidt vector ${\underline{\lambda}}$.
Hence, by Lemma \ref{lem},  
$T_3(\underline{\lambda})$ is a monotonically non--increasing function of the
elementary symmetric polynomials 
\begin{equation}
\s_{k}({\underline{\lambda }}):=s_{k}(\gamma _{1},\gamma _{2},\ldots ,\gamma
_{d}),\quad k=0,\ldots ,d.  \label{wides}
\end{equation}
Therefore, to prove Theorem \ref{theo2} it suffices to show 
that the functions $\s_{k}({\underline{\lambda }})$ are Schur concave in 
$\underline{\lambda }\in \Sigma _{d}$. This Schur--concavity property
of $\s_{k}({\underline{\lambda }})$ was proved explicitly in  
\cite{suff}. The proof of Lemma \ref{lem} therefore allows us to establish
Theorem \ref{theo2}, and hence Theorem \ref{theo}, our main result. 
This is given in 
Section \ref{proof}. Our proof is analogous to that of Theorem 1 of
\cite{JM}. 

\section{Proof of Lemma \ref{lem}}
\label{proof}
The variables $x_1, x_2, \ldots, x_n$, defined in Lemma \ref{lem}, 
can be viewed as the eigenvalues of 
an $n \times n$ density matrix $\rho_n$ (say), and hence
as the roots of the characteristic equation ${\hbox{det}} 
\bigl(\rho_n - x {\iden} \bigr)=0$. Since the roots are the zeros of the
product $\prod_{i=1}^n (x - x_i)$, the characteristic equation can
be expressed in terms of these roots as follows:
\begin{equation}
\sum_{k=0}^{n}x ^{k}\,(-1)^{n-k}\,s_{n-k}(x_1, x_2, \ldots, x_n)=0.  
\label{sym1}
\end{equation}
Here the coefficient $s_{n-k}(x_1, x_2, \ldots, x_n)$ denotes the 
$(n-k)^{th}$ elementary symmetric polynomial of the variables $x_1, x_2, \ldots, x_n$
(defined by \reff{sym}). 
We consider equation \reff{sym1} to implicitly define the variables
$x_j \equiv x_j(s_1, s_2, \ldots, s_n)$ as functions of the elementary symmetric
polynomials. 
This can be done unambiguously as long as there are
no multiple roots, i.e. $x_i \ne x_j$ for $i \ne j$, $i,j = 1,2,\ldots, n$
We restrict our attention to this case at first, 
and prove that {\em{in the absence of
multiple roots}}, ${\displaystyle{\partial f/\partial s_i} < 0}$ for 
each $i \ge 2$. This will enable us, by continuity arguments, to conclude
that $f$ is indeed a monotonically non--increasing function 
of the elementary symmetric polynomials $s_2, s_3, \ldots, s_n$ everywhere.

Let us first prove that ${\displaystyle{\partial f/\partial s_i} < 0}$
when the roots $x_1, x_2, \ldots, x_n$ are all different. In this case
we can view the variables $x_j$ to be implicitly defined by \reff{sym1}. 
Then differentiating with respect to $s_m$, for $2\le m\le n$, 
we get
\be
\frac{\partial x_j}{\partial s_m} = \frac{(-1)^{m+1} x_j^{n-m}}{\prod_{i\ne j}
(x_j - x_i)}.
\ee
Using the chain rule and the definition of the function $f(\x)$ we get
\bea
\frac{\partial f}{\partial s_m} &=& \sum_{j=1}^n 
\frac{\partial f}{\partial x_j}
\frac{\partial x_j}{\partial s_m}\nonumber\\
&=& \sum_{j=1}^n
\frac{(-1)^{m+1} x_j^{n-m}\, p x_j^{p-1} }{\prod_{i\ne j}
(x_j - x_i)}.
\label{der}
\eea

To prove that $f(\x)$ is a monotonically non--increasing
function of $s_m$ for each $m =2,3, \ldots, n$, we use some 
standard results from Numerical Analysis \cite{num2}. 
It is known that there is a unique polynomial 
of degree $(n-1)$ which interpolates a given function $g(x)$
at the points $x_1, x_2, \ldots, x_n$.
The coefficient of 
$x^{n-1}$ of this polynomial is given by
$$a_{n-1} =  \sum_{j=1}^n
\frac{g(x_i)}{\prod_{i\ne j}
(x_j - x_i)},$$
called the Newton Divided Difference \cite{num2} 
of the function $g(x)$. 
The expression on the 
RHS of \reff{der} implies that 
${{\partial f}/{\partial s_m}}$ 
is the Newton Divided difference of the following function
\be
g(x) \equiv g_m(x) =  (-1)^{m+1} p\,x^{n-m+p-1}.
\label{gee}
\ee
By the Hermite Gennochi theorem \cite{num2}, the Newton Divided Difference
is also given by the integral of $g^{(n-1)}(p_1 x_1 + \ldots + p_n x_n)$
over the probability simplex $\{(p_1, p_2, \ldots, p_n), p_i \ge 0, 
\sum_{i=1}^n p_i =1\}$, where $g^{(n-1)}(x)$ denotes the $(n-1)^{th}$
derivative of $g(x)$. 

From \reff{gee} we obtain
\be
g_m^{(n-1)}(x) = (-1)^{m+1} p (p-m+1) (p-m+2)\ldots (p-m+n-1)\, x^{p-m+1}.
\ee
It is easy to see that for all $2\le m \le n$, $g_m^{(n-1)}(x) < 0$ for
all $x$,  
since $1 < p < 2$. Hence the integral over the probability simplex
is negative and we get 
${{{\partial f}/{\partial s_m} < 0}}$ as
required.

Therefore, ${{{\partial f}/{\partial s_m} < 0}}$ everywhere
except on the manifolds on which two or more of the roots $x_1, x_2, \ldots,
x_n$ coincide. By continuity we deduce that $f$ is a monotonically 
non--increasing function of the
elementary symmetric polynomials $s_2, s_3, \ldots, s_n$ everywhere.
\qed


Note that in Lemma \ref{lem} we considered $1 <p <2$. For the case $p=2$,
proceeding analogously to the proof of the above lemma, we find the 
following: {\em{in the absence of multiple roots}}
\be
\frac{\partial f}{\partial s_2} < 0 \quad {\hbox{whereas }} 
\quad \frac{\partial f}{\partial s_m} = 0 \,\, {\hbox{for all }} m=3,4, \ldots, n.
\ee
Hence by continuity, $f$ is a monotonically 
non--increasing function of the
elementary symmetric polynomial $s_2$ everywhere. 
The latter is however a Schur--concave
function of the Schmidt vector ${\underline{\lambda}}$ \cite{suff}. Hence
for the case $p=2$ as well, Theorem \ref{theo2} applies and the 
multiplicativity stated in Theorem \ref{theo} holds.

\section*{Appendix}
A real--valued function $\Phi $ on ${\mathbf{R}}^{n}$ is said to be \emph{
Schur convex} (see \cite{bha}) if
\begin{equation*}
{\underline{x}}\prec {\underline{y}}\quad \implies \Phi ({\underline{x}}
)\leq \Phi ({\underline{y}}).
\end{equation*}
Here the symbol ${\underline{x}}\prec {\underline{y}}$ means that ${\
\underline{x}}=(x_{1},x_{2},\ldots ,x_{n})$ is \emph{majorized} by ${\
\underline{y}}=(y_{1},y_{2},\ldots ,y_{n})$ in the following sense: Let ${
\underline{x}}^\downarrow$ be the vector obtained by rearranging the
coordinates of ${\underline{x}}$ in decreasing order
\begin{equation*}
{\underline{x}}^\downarrow = (x_1^\downarrow, x_2^\downarrow, \ldots,
x_n^\downarrow)\quad {\hbox{means   }} x_1^\downarrow \ge x_2^\downarrow \ge
\ldots \ge x_n^\downarrow .
\end{equation*}
For ${\underline{x}}, {\underline{y}} \in {\mathbf{R}}^n$, we say that ${\
\underline{x}}$ is majorized by ${\underline{y}}$ and write ${\underline{x}}
\prec {\underline{y}}$ if
\begin{equation*}
\sum_{j=1}^k x_{j}^\downarrow \le \sum_{j=1}^k y_{j}^\downarrow, \quad 1\le
k \le n,
\end{equation*}
and
\begin{equation*}
\sum_{j=1}^n x_{j}^\downarrow = \sum_{j=1}^n y_{j}^\downarrow.
\end{equation*}
In the simplex $\Sigma_d$, defined by the constraints (\ref{cons}), the
minimal point is $(1/d, \ldots, 1/d)$ (the baricenter of $\Sigma_d$), and
the maximal points are the permutations of $(1,0, \ldots,0)$ (the vertices).

\section{Acknowledgements} 
N.D. acknowledges R.F. Werner for suggesting the problem and 
for offering a patient ear, to Y.M. Suhov and
G. Mitchison for helpful discussions and finally to
M.B. Ruskai for highlighting the recent work in \cite{af}.

\end{document}